\documentclass[10pt,conference]{IEEEtran}
\IEEEoverridecommandlockouts
\usepackage{cite}
\usepackage{url} % for \url in IEEEtran.bst
\usepackage{amsmath,amssymb,amsfonts,amsthm}
\usepackage[ruled,vlined,linesnumbered]{algorithm2e}
\usepackage{graphicx}
\usepackage{textcomp}
\usepackage{xcolor}
\definecolor{agolicCoverageHighlight}{HTML}{F1E5FF}
\definecolor{agolicModeHighlight}{HTML}{E4F0FF}
\definecolor{agolicWitnessHighlight}{HTML}{FFF0D5}
\newcommand{\coveragehl}[1]{\colorbox{agolicCoverageHighlight}{#1}}
\newcommand{\modehl}[1]{\colorbox{agolicModeHighlight}{#1}}
\newcommand{\witnesshl}[1]{\colorbox{agolicWitnessHighlight}{#1}}
\usepackage{xspace}
\usepackage{pgfplots}
\newcommand{\agolic}{\textsc{Agolic}\xspace}
\SetKwInOut{Input}{Input}
\SetKwInOut{Output}{Output}

\theoremstyle{definition}

\usepackage[hidelinks]{hyperref}
\def\BibTeX{{\rm B\kern-.05em{\sc i\kern-.025em b}\kern-.08em
    T\kern-.1667em\lower.7ex\hbox{E}\kern-.125emX}}
\begin{document}

\title{\agolic: Agentic Planning for Symbolic Execution%
\thanks{The implementation is available from the authors upon request.}}

\author{%
\IEEEauthorblockN{%
Daniel Koh\IEEEauthorrefmark{1},
Yannic Noller\IEEEauthorrefmark{2},
Corina S. P\u{a}s\u{a}reanu\IEEEauthorrefmark{3}, and
Youcheng Sun\IEEEauthorrefmark{1}}
\IEEEauthorblockA{\IEEEauthorrefmark{1}Mohamed bin Zayed University of Artificial Intelligence,
Abu Dhabi, United Arab Emirates\\
\texttt{\{daniel.koh,youcheng.sun\}@mbzuai.ac.ae}}
\IEEEauthorblockA{\IEEEauthorrefmark{2}Ruhr-Universit\"at Bochum,
Bochum, Germany\\
\texttt{yannic.noller@ieee.org}}
\IEEEauthorblockA{\IEEEauthorrefmark{3}Carnegie Mellon University,
Pittsburgh, PA, USA\\
\texttt{pcorina@andrew.cmu.edu}}}
\maketitle

\begin{abstract}
Symbolic execution seeks to explore feasible program paths, yet a practical run may exhaust its resources while much program behaviour remains unreached. We investigate a complementary way of extending its practical reach by reasoning about how the same tool is utilised from one bounded run to the next, while leaving ordinary state exploration to the underlying tool.

We present \agolic, an agentic planning system that uses evidence from earlier runs to choose and configure later bounded symbolic execution (BSE) runs, which the underlying symbolic execution tool then carries out. The planning intelligence, available evidence and execution modes can be adapted to the symbolic execution tool and analysis objective. We evaluate one adaptation for branch-coverage exploration, in which an LLM-based agent reasons over source code, replayed coverage and earlier targeting attempts.

We evaluate \agolic on several C and C++ programs. On every program, it extends the branch coverage obtained by continuous symbolic execution and covers more than $3\times$ as many branches on average. It also covers more branches than each individual corpus from coverage-guided fuzzing and compiler-based concolic execution in our evaluation and reaches branches absent from all comparison corpora combined on six of the seven programs. Taken together, these results point to considerable untapped potential in existing symbolic execution tools, some of which may be realised by reasoning about how their capabilities are used across runs while leaving state selection during ordinary symbolic exploration to the underlying tool.

\end{abstract}

\section{Introduction}
\label{sec:introduction}

Program analysis is a well-established means of studying program behaviour and plays an important role in reasoning about the reliability and security of software. Symbolic execution \cite{king1976symbolic} has the appealing property that one symbolic input can represent a family of concrete inputs. It seeks to explore every feasible execution path within a chosen bound, recording along each one a path condition that characterises the inputs under which it can be taken. For a satisfiable path condition, a tool that supports test generation can obtain concrete input values from a constraint solver and encode them as a concrete test.

Yet this strength comes at a cost. At an input-dependent branch, more than one execution path may be feasible, and their number can grow exponentially as such branches accumulate. Such growth can quickly become intractable for a bounded symbolic-execution run, while difficult constraint queries may consume much of the available time \cite{cadar2013symbolic,baldoni2018survey}. A run may consequently exhaust its time or memory while other feasible program behaviour remains unreached.

Beginning another bounded run does not by itself tell us why feasible program behaviour remained unreached or what should change. If we can reason about the program and the evidence produced by its earlier bounded runs, we may be able to intervene more usefully by choosing and configuring the next run differently. We therefore study how such reasoning can help us make more intelligent use of existing symbolic execution capabilities while the tool retains control of ordinary state exploration within each run.

We already know that symbolic execution can benefit from information outside its current search. Seeded and hybrid approaches use concrete executions to guide later symbolic work or connect symbolic exploration with fuzzing \cite{stephens2016driller,noller2018badger}. Pending constraints defer branch-feasibility checks until they are needed, while memoised execution records paths so that they can be recovered later \cite{kapus2020pending,busse2020running}. Adaptive configuration changes external parameters over repeated invocations \cite{cha2022symtuner}, and recent agent-based systems perform concolic reasoning, propose concrete inputs or synthesise harnesses for selected code \cite{luo2026agentic,chen2026concolixir,shafiuzzaman2026guidingsymbolicexecutionstatic}.

In this paper we present \agolic, an agentic planning system that places a planner between bounded symbolic-execution (BSE) runs carried out under assigned resource budgets. The planner uses evidence from earlier runs of the same program to choose and configure later runs, which the underlying symbolic execution tool then carries out. We design \agolic so that its planning intelligence, evidence and execution modes can be chosen according to the symbolic execution tool and analysis objective (e.g., code coverage and bug finding). We present one realisation of this design for branch-coverage analysis, in which an LLM-based agent uses source code, replayed coverage and the history of earlier targeting attempts. Alongside a continuous run of symbolic execution, the planner dispatches BSE runs over successive planning rounds.

To carry out these plans in the present study, we provide two execution modes, although the architecture is not limited to them. In \emph{Harness-Entry Mode}, ordinary symbolic exploration begins at the harness. In \emph{Witness-Guided Mode}, execution under a concrete witness establishes a program state from which ordinary symbolic exploration may resume. We use native replay of generated tests and materialised witnesses to supply coverage feedback for later planning and to determine the branch sets reported in the evaluation.

We use this design to investigate two questions concerning \agolic's reach and the role of reasoning outside the symbolic execution tool.

\begin{description}
  \item[\textbf{RQ1}] How does the branch coverage obtained by \agolic compare with that obtained by continuous symbolic execution and the other evaluated program-analysis systems?
  \item[\textbf{RQ2}] How does \agolic's planning loop use evidence from earlier runs when configuring later ones?
\end{description}

We evaluate \agolic on several C and C++ programs. The evaluation shows that \agolic extends the branch coverage obtained by continuous symbolic execution on every program and covers more than $3\times$ as many branches on average. On every program, \agolic also covers more branches than every evaluated corpus from coverage-guided fuzzing and compiler-based concolic execution. Compared with ConcoLLMic, \agolic exceeds all three runs on five programs and reaches branches absent from all three ConcoLLMic runs on every program.

We contribute an architecture for planning between runs, in which a planner uses information from the program and earlier runs to choose and configure target-specific BSE runs that the underlying symbolic execution tool carries out. We realise this architecture with Harness-Entry Mode, Witness-Guided Mode, target admission and native-replay validation of generated tests. We evaluate its branch totals and identities across several programs and record planning activity, witness use and resource accounting.

\section{Background and Related Work}
\label{sec:background}

\subsection{Symbolic Execution under Resource Bounds}

Symbolic execution was introduced as a way of reasoning about program executions over symbolic inputs \cite{king1976symbolic}. Later work showed that it could achieve high coverage on real systems programs \cite{cadar2008klee}. In practice, finite time and memory still leave feasible paths unexplored as paths multiply and constraint-solving costs accumulate \cite{cadar2013symbolic,baldoni2018survey}. The approaches below respond to this practical limit by changing what is explored within a run, connecting symbolic execution with other techniques or adapting successive invocations.

Some approaches make unfinished symbolic work available later. Pending constraints delay selected feasibility checks and prioritise paths already known to be feasible, while memoised symbolic execution records ongoing exploration so that selected paths can be restored in later executions \cite{kapus2020pending,yang2012memoized,busse2020running}. These techniques show the value of carrying unfinished work forward. We ask instead whether evidence from completed runs can inform how a new bounded run is configured.

\subsection{Directed, Targeted and Hybrid Testing}

One response to saturation is to steer symbolic execution towards underexplored regions. Directed Symbolic Execution \cite{ma2011directed} formalises this as a shortest-distance metric over the interprocedural control-flow graph; statistical approaches bias expansion towards less-frequented path spectra \cite{li2013steering}. At the function level, Under-Constrained Symbolic Execution \cite{ramos2015underconstrained} and Chopped Symbolic Execution \cite{trabish2018chopped} bypass full program context and exclude uninteresting code respectively; Munch \cite{ognawala2018munch} and directed greybox fuzzing \cite{bohme2017directed} carry the same goal-directed intuition into hybrid and mutation-based settings.

Concolic execution \cite{godefroid2005dart,sen2005cute,godefroid2008automated} drives exploration by negating branch conditions along a concrete trace. Driller \cite{stephens2016driller} pairs this with coverage-guided mutation, employing each technique where the other stalls. Badger runs fuzzing and bounded symbolic execution in tandem and exchanges inputs between its AFL-based fuzzer and Symbolic PathFinder to search for worst-case executions \cite{noller2018badger}. SymCC \cite{poeplau2020symcc} and SymQEMU \cite{poeplau2021symqemu} compile concolic instrumentation directly into the target binary, substantially reducing the per-instruction overhead of interpretation-based approaches.

Concrete inputs and additional workers offer two other ways of extending a symbolic search. Seeded execution uses supplied inputs to prioritise paths known to be feasible, while ZESTI begins bounded symbolic exploration at selected divergence points around sensitive operations \cite{kapus2020pending,marinescu2012zesti}. Parallel systems address the same resource limit by distributing a symbolic execution across workers, as Cloud9 does, or by partitioning the search through static preconditions or independently executable ranges \cite{bucur2011cloud9,staats2010parallel,siddiqui2012ranged}. \agolic treats neither a seed nor a partition of one search as its unit of control. It uses evidence accumulated across completed runs to choose a later run's specification, including its invocation and symbolic input surface, together with a witness and release function where the present configuration calls for them.

\subsection{AI-Guided Program Analysis}
\label{sec:background-ai}

Learned search policies have been applied to coverage estimation and state pruning \cite{he2021learning,cha2019enhancingdynamicsymbolicexecution,cha2020making}, and neural methods have been applied to gradient-guided input generation for fuzzing \cite{she2019neuzzefficientfuzzingneural}. SymTuner tunes external parameters across successive symbolic-execution invocations rather than modifying the search within a run, and provides a close prior analogue to the external control architecture employed here \cite{cha2022symtuner}.

Language models have been applied to symbolic path selection in smart contracts \cite{so2021smartest} and to LLM-driven constraint solving for structured inputs \cite{tu2025cottontaillargelanguagemodeldriven}, and their reach extends to universal input generation across diverse target languages \cite{xia2024fuzz4all}. ConcoLixir uses an LLM to propose concrete inputs when concolic exploration stalls or constraint solving fails \cite{chen2026concolixir}. SAILOR combines static analysis with iterative LLM synthesis of the drivers, stubs and assertions needed for symbolic execution, and validates the resulting tests through concrete replay \cite{shafiuzzaman2026guidingsymbolicexecutionstatic}. ConcoLLMic \cite{luo2026agentic} places an LLM agent inside a concolic execution loop; the agent analyses concrete execution traces and constructs path constraints to reach branches not yet taken, with coverage feedback informing the direction of exploration.

It is useful here to adopt the distinction made by Cha et al. between online and offline symbolic execution \cite{cha2012mayhem}. Online symbolic execution explores paths within one run by forking execution states, and a search policy determines which pending state to advance. Offline, or trace-based concolic, execution follows one concrete trace at a time and uses the collected path constraints to derive later inputs. Recent agent-guided concolic systems place a model within this trace-level loop, whereas \agolic takes each bounded symbolic-execution run as its unit of planning. Between runs, its planner uses evidence produced so far to configure what the symbolic execution tool should attempt next, while state selection during ordinary symbolic exploration remains with the tool. Concrete values may be used to establish an execution prefix without changing this division of work, since ordinary symbolic exploration resumes at the selected release point. In the branch-coverage configuration evaluated here, the evidence consists of replay-derived source-level coverage and the record of previous targeting attempts.

\section{Agolic: Run-Level Planning for Symbolic Execution}
\label{sec:approach}

\agolic rests on a simple division of labour in which a planner uses evidence about the target program and earlier runs to configure later bounded symbolic-execution runs, while the underlying symbolic execution tool carries out each run. We take the bounded run as the unit of control and allow the planning objective, evidence and available run configurations to vary with the application. We describe the tool-specific implementation choices in \autoref{sec:experimental-setup}.

\begin{figure*}[!t]
\centering
\includegraphics[width=\textwidth]{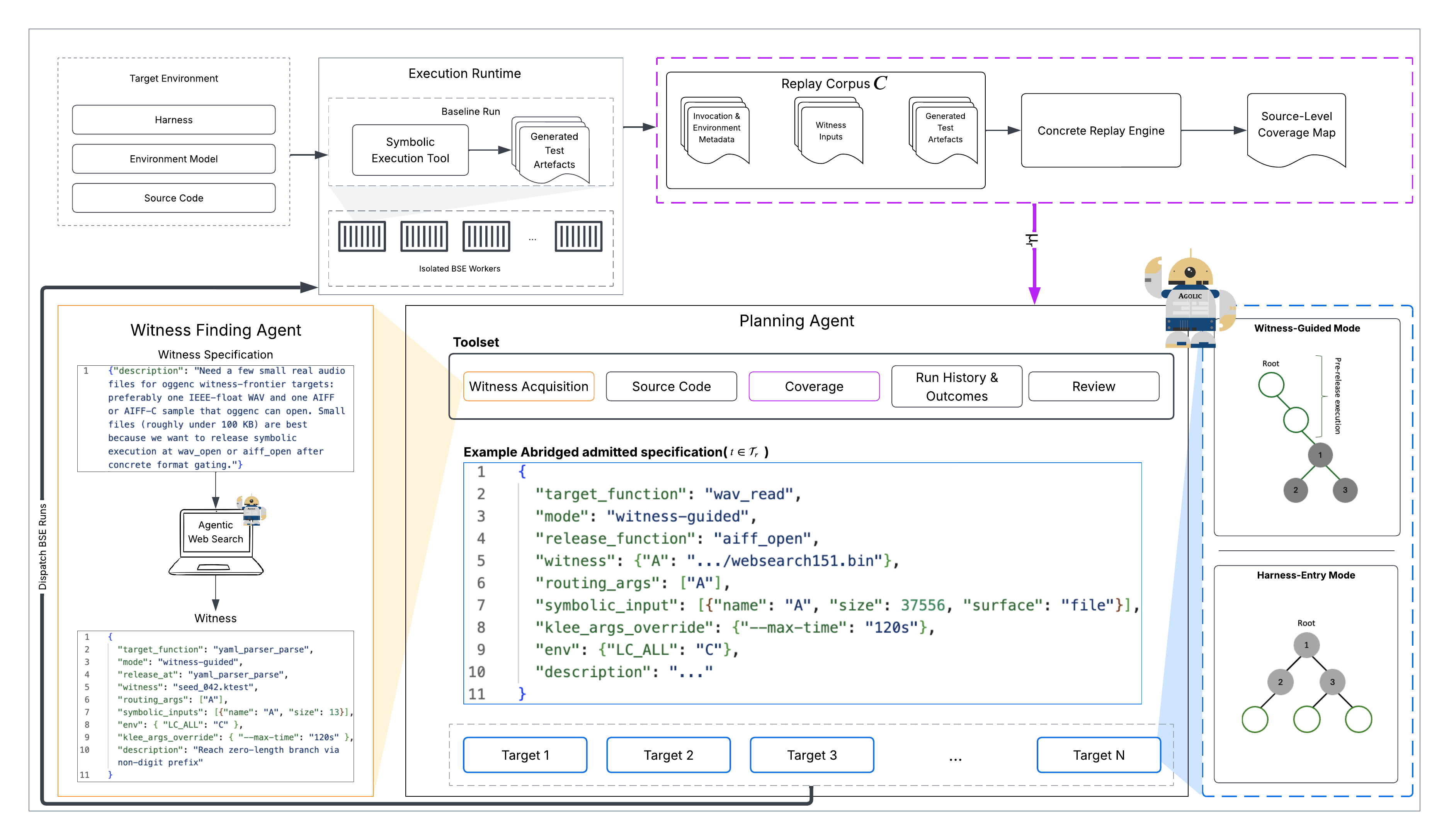}
\caption{The \agolic \coveragehl{branch-coverage configuration} evaluated in this paper. The planner selects among the \modehl{available BSE modes} and may request \witnesshl{witness acquisition} when a suitable input is not already present. Generated test artefacts and materialised witnesses selected by the planner enter the round-level replay corpus. Concrete replay of this corpus supplies source-level coverage, and recorded run outcomes inform later planning. The final replay described in \autoref{sec:experiment-measurement} determines the reported branch sets. Algorithm~\ref{alg:agolic} sets out the round-level procedure.}
\label{fig:architecture}
\end{figure*}

The \coveragehl{branch-coverage configuration} evaluated in this paper is shown in \autoref{fig:architecture}. A continuous symbolic-execution run begins at the target program's test-harness entry point and contributes generated tests alongside the planner-dispatched runs. At each planning round, the agent examines replay-derived source-level coverage and the record of earlier BSE attempts. It uses this evidence to identify program regions that remain underexplored and to formulate new target specifications. Admitted specifications are dispatched to isolated BSE workers under their assigned time and memory bounds.

\subsection{Round-Level Procedure}
\label{sec:orchestration}

We divide the execution budget into planning rounds. Before the first round, \textsc{StartContinuousRun} begins the continuous symbolic-execution run, which remains active alongside the round-level procedure (line~\ref{alg:line-start-continuous}). At the start of round $r$, \textsc{ReplayCoverage} collects test artefacts completed by the continuous run since the preceding replay and concretely replays the current corpus $C$ to derive the source-level coverage map $\mu_r$ (line~\ref{alg:line-replay-coverage}). It gives this map and the execution history $H$ to the planning agent. Execution stops at the deadline, when too little time remains for another planning and BSE round or when the planner produces no new admitted target. An unsuccessful planning session is recorded and retried while time remains, whereas the empty-set condition in Algorithm~\ref{alg:agolic} applies to a completed session that produces no admitted target. The \textsc{CanStartRound} test below enforces the time budget.

Algorithm~\ref{alg:agolic} summarises the procedure used in this configuration. The planning agent uses the source code, coverage map and history to formulate the proposed target specifications $Q_r$ (line~\ref{alg:line-plan}). Review yields the admitted targets $\mathcal{T}_r$ (line~\ref{alg:line-admit}), whose members are dispatched as BSE runs (line~\ref{alg:line-execute}). As each run completes, \textsc{ReplayAndRecord} adds its generated artefacts and any materialised witness to the replay corpus and its outcome to the execution history (line~\ref{alg:line-replay-record}). After the loop, \textsc{FinaliseContinuousRun} stops continuous execution and adds any remaining tests to the final round-level corpus (line~\ref{alg:line-finalise-continuous}).

\begin{algorithm}[!t]
\caption{Round-level procedure for the \coveragehl{branch-coverage configuration} of \agolic.}
\label{alg:agolic}
\Input{target program $P$, wall-clock budget $B$, planning agent $M$}
\Output{final replay corpus $C$ and execution history $H$}

$C \leftarrow \emptyset$;
$H \leftarrow \emptyset$;
$r \leftarrow 0$\;

\textsc{StartContinuousRun}$(P)$\nllabel{alg:line-start-continuous}\;

\While{\textsc{CanStartRound}$(B, H)$}{
    $r \leftarrow r + 1$\;
    $(C,\mu_r) \leftarrow{}$ \coveragehl{\textsc{ReplayCoverage}}$(P, C)$\nllabel{alg:line-replay-coverage}\;
    $Q_r \leftarrow \textsc{Plan}(M, P,\mu_r, H)$\nllabel{alg:line-plan}\;
    $\mathcal{T}_r \leftarrow \textsc{AdmitTargets}(P, Q_r, H)$\nllabel{alg:line-admit}\;
    \If{$\mathcal{T}_r = \emptyset$}{\textbf{break}\;}
    \ForEach{$R_i \in \textsc{ExecuteBSERuns}(P, \mathcal{T}_r)$\nllabel{alg:line-execute}}{
        $(C, H) \leftarrow{}$ \coveragehl{\textsc{ReplayAndRecord}}$(P, C, H, R_i)$\nllabel{alg:line-replay-record}\;
    }
}
$(C, H) \leftarrow{}$ \coveragehl{\textsc{FinaliseContinuousRun}}$(P, C, H)$\nllabel{alg:line-finalise-continuous}\;

\Return{$C, H$}\;
\end{algorithm}

\subsection{Evidence and Feedback Between Runs}
\label{sec:coverage}

The evidence made available to the planner may vary with the analysis objective. In the branch-coverage configuration studied here, we derive planning feedback from replayed source-level coverage during the \textsc{ReplayCoverage} operation on line~\ref{alg:line-replay-coverage} of Algorithm~\ref{alg:agolic}. Each generated test artefact and materialised witness is submitted to concrete replay using its associated environment. Completed replays record both the exercised source elements and the remaining uncovered branches and lines.

To define this feedback mathematically, let $P$ be the target program, let $E$ be its set of source-level coverage elements and let $\mathcal{A}$ be the universe of replayable input artefacts for $P$. Strictly speaking, $\mathcal{A}$ is infinite, as it encompasses all possible input sequences of arbitrary length. Any replay corpus maintained by the planner, however, is fundamentally finite. We therefore restrict our attention to finite subsets and define the coverage mapping $\mu: \mathcal{P}_{\mathrm{fin}}(\mathcal{A}) \to \mathcal{P}(E)$ such that for any finite corpus $C \subseteq \mathcal{A}$, $\mu(C) \subseteq E$ yields the subset of elements exercised by completed concrete replays. The corpus accumulates generated test artefacts and materialised witnesses selected by the planner. At the start of round $r$, $\mu_r$ denotes the coverage map supplied to the planning agent for that round. Given a prior corpus $C$ and a BSE run $i$ yielding generated artefacts $A_i \subseteq \mathcal{A}$, the coverage delta $\Delta_i$ is defined as the set difference:
\begin{equation}
  \Delta_i = \mu(C \cup A_i) \setminus \mu(C).
  \label{eq:coverage-delta}
\end{equation}
For run $i$, $C$ is the corpus available immediately before $A_i$ is recorded. It may therefore contain artefacts produced by the continuous run and earlier BSE runs since $Q_r$ was planned, together with materialised witnesses selected by the planner. A witness associated with run $i$ enters $C$ before $A_i$ is evaluated, so the set difference attributes to $A_i$ only the additional coverage supplied by the generated artefacts. Since completed runs are replayed and recorded one at a time, it also prevents the double-counting of coverage elements already present at that point. The final measurement applies the replay procedure in \autoref{sec:experiment-measurement} to the resulting corpus.

To record whether a BSE run entered its target function and added branch coverage there, we collect the replay evidence detailed in \autoref{tab:target-evidence}. These classifications describe only what happened under the concrete-replay procedure in \autoref{sec:experiment-measurement}. A target classified as \textsc{Not reached} may still be reachable.

\begin{table}[!b]
\renewcommand{\arraystretch}{1.15}
\caption{Replay-derived target evidence.}
\label{tab:target-evidence}
\centering
\footnotesize
\begin{tabular}{@{}p{0.31\columnwidth}p{0.61\columnwidth}@{}}
\hline
\textbf{Evidence class} & \textbf{Operational criterion}\\
\hline
\textsc{New reach} &
Replay of $A_i$ enters the target function, which is not entered by replay of the prior corpus $C$.\\
\textsc{Increased target coverage} &
Replay of $A_i$ enters the target function already entered by replay of $C$, and $\Delta_i$ contains at least one additional branch within that function.\\
\textsc{Reached, no gain} &
Replay of $A_i$ enters the target function already entered by replay of $C$, and $\Delta_i$ contains no additional branch within that function.\\
\textsc{Not reached} &
Replay of $A_i$ does not enter the target function.\\
\hline
\end{tabular}
\end{table}

The \textsc{ReplayAndRecord} operation on line~\ref{alg:line-replay-record} adds any materialised witness and generated artefacts to $C$ and records the resulting outcome in $H$, making both available to later \textsc{Plan} operations.

\subsection{Target Planning and Validation}
\label{sec:planner}

For the \textsc{Plan} operation on line~\ref{alg:line-plan} of Algorithm~\ref{alg:agolic}, we supply the planning agent with program facts, a high-level summary of $\mu_r$ and the current execution history $H$. The detailed coverage map, source code and prior BSE run outcomes remain available through \agolic's planner-facing interface, whose evaluated operations are summarised in \autoref{tab:tools}. The agent inspects the source and $\mu_r$ to identify possible coverage targets, then compares related BSE configurations and their recorded outcomes in $H$ before drafting $Q_r$.

We refer to each $q \in Q_r$ as a proposed target specification. After review, an admitted specification is denoted by $t \in \mathcal{T}_r$. The target function identifies the program region that the run is intended to explore, while the routing arguments, symbolic inputs and optional environment bindings define the invocation supplied to the symbolic execution tool. The specification also records the intended coverage opportunity and, when related attempts exist, how the proposed configuration differs from those recorded in $H$.

For witness-guided BSE, the specification additionally supplies a concrete value for each declared symbolic input surface and names a release function. We refer to this concrete input assignment as the witness and to its stored replayable form as a materialised witness. The release function may precede the target function along the intended path. The planner may select a witness already available in the replay corpus or workspace. When neither contains a suitable input, \witnesshl{optional witness acquisition} may supply candidates during \textsc{Plan}.

The \textsc{AdmitTargets} operation on line~\ref{alg:line-admit} of Algorithm~\ref{alg:agolic} reviews each proposal before it enters $\mathcal{T}_r$. A proposal that cannot be carried out as specified is rejected, and the resulting diagnostics allow the planner to revise $Q_r$ before requesting another review. An admitted proposal enters $\mathcal{T}_r$ as $t$ and is passed to \textsc{ExecuteBSERuns}.

\subsection{BSE Execution Modes}
\label{sec:bse-modes}

We cannot fairly assume that every symbolic execution tool organises exploration in the same way or supports the same execution modes. In what follows, we concern ourselves with the online form described in \autoref{sec:background-ai}. Since the modes made available to the planner depend on the symbolic execution tool and the analysis objective at hand, the \agolic architecture leaves them to each configuration.

For the \coveragehl{branch-coverage configuration} shown in \autoref{fig:architecture}, we provide two modes. Each $t \in \mathcal{T}_r$ passed to \textsc{ExecuteBSERuns} on line~\ref{alg:line-execute} of Algorithm~\ref{alg:agolic} is executed from the benchmark harness under the admitted invocation. \modehl{Harness-Entry Mode} is the plainer of the two, with ordinary symbolic exploration proceeding from the initial harness state under the configured search policy. In \modehl{Witness-Guided Mode}, the tool uses a witness to advance a single state towards the release function. At the first encounter with the named function, the state is released and ordinary symbolic exploration resumes from it.

We use Harness-Entry Mode when decisions before the target function form part of the coverage opportunity or when the planner has no available input known to reach a useful release function. Witness-Guided Mode is intended for program regions whose setup is expensive to explore symbolically. The planner selects this mode by supplying a witness and release function in $q$, which it may revise in a later round after considering $\mu_r$ and related evidence in $H$.

\subsection{Witness-Guided BSE}
\label{sec:witness-guided-bse}

Consider an admitted $t \in \mathcal{T}_r$ in Witness-Guided Mode. During pre-release execution, the symbolic execution tool uses the witness to resolve supported input-dependent operations. The resulting values select branch outcomes, determine allocation sizes and resolve symbolic addresses. For these operations, the witness supplies the required values in place of the corresponding constraint-solver queries and execution advances without forking the state. The tool adds route constraints when it follows witness-selected decisions or records concrete effects, but it does not encode every dependence of the pre-release execution. As execution proceeds from the harness, program effects update the single state. If a required value cannot be obtained from the witness, the pre-release state terminates. No test artefact is generated from a pre-release state that fails to reach the release function within the run budget.

At the release-function boundary, ordinary solver-backed symbolic exploration resumes from the reached state under the search policy configured for the run. This state retains the memory, path condition and calling context established during execution from the harness. Post-release exploration is restricted to paths consistent with the retained state.

Witness-guided BSE is a narrow instrument whose usefulness depends on where execution under the witness gives way to ordinary symbolic exploration. If the aim is to explore beyond the witness-selected path, useful input-dependent choices must remain after the release function. An early release leaves more of the setup path to ordinary symbolic exploration. A late release may preserve decisions that already determine the target branch. There is equally little to gain when replay has already saturated the code reached after release. Although reaching the release function establishes one execution path to it under the admitted invocation, it does not \emph{ipso facto} determine whether post-release exploration reaches the target function or whether $\Delta_i$ contains a new branch there. We return to this release-boundary trade-off in \autoref{sec:discussion-concrete-prefix}.

After release, the symbolic execution tool may generate a test artefact from an assignment satisfying the state's accumulated path condition. If the state retains a pre-release effect whose dependence on the witness is absent from the path condition, replaying the assignment from the original harness may still follow a different execution. Generated artefacts are therefore submitted to concrete replay, and only behaviour observed during replay contributes to $\Delta_i$ and the target classifications in \autoref{sec:coverage}. The concrete assignment policy used by our implementation is given in \autoref{sec:experimental-setup}.

\section{Experiment}
\label{sec:experiment}

\subsection{Implementation}
\label{sec:experimental-setup}

We implement \agolic as a Python runtime around KLEE and add an LLVM 16 extension for Witness-Guided Mode \cite{cadar2008klee}. The runtime coordinates planning, target review, worker dispatch and concrete replay. Its planning agent invokes the operations in \autoref{tab:tools} through the OpenAI Agents SDK \cite{openaiAgentsSDK}.

\begin{table}[!t]
\renewcommand{\arraystretch}{1.18}
\caption{Planner-facing tools in the evaluated implementation. The final two rows provide the objective- and mode-specific operations used in this configuration.}
\label{tab:tools}
\centering
\footnotesize
\begin{tabular}{@{}p{0.32\columnwidth}p{0.60\columnwidth}@{}}
\hline
\textbf{Tool class} & \textbf{Representative operations}\\
\hline
Run history and outcomes &
\textsc{read\_planner\_history}, \textsc{read\_bse\_run\_detail}, \textsc{read\_campaign\_summary} and \textsc{read\_round\_trace} expose prior attempts, replay status, target evidence, BSE-run coverage deltas, symbolic-execution statistics and planning trace events.\\
Source and workspace inspection &
\textsc{read\_source\_range\_tool} exposes numbered source context, while \textsc{run\_workspace\_command} provides bounded read-only workspace inspection through an allowlist of Unix-style commands.\\
Target proposal and review &
\textsc{apply\_patch}, \textsc{review\_targets\_file} and \textsc{generate\_targets} materialise $Q_r$, return review diagnostics and submit the unchanged reviewed specifications from which $\mathcal{T}_r$ is formed.\\
\hline
\coveragehl{Coverage evidence} &
\textsc{list\_coverage\_status} and \textsc{read\_coverage\_context} expose branch frontiers, unreached branches, partial functions and uncovered functions.\\
\witnesshl{Witness acquisition} &
When enabled, \textsc{request\_witness\_candidates} asks the witness-finding agent for candidate inputs and returns paths, sizes, content hashes, source kinds and provenance.\\
\hline
\end{tabular}
\end{table}

When the planner requests witness candidates, a dedicated witness-finding agent searches for and retrieves public input files from the web. The planning agent may then associate a selected candidate with the symbolic input it supplies.

We admit a proposed target specification only after resolving its target function and source file and checking its argument structure, symbolic input surfaces, environment bindings and concrete input assignments. Duplicate harness-entry invocations within $Q_r$ are rejected. Proposals that pass these checks undergo a bounded preflight execution to expose missing artefacts, invalid arguments and shallow program failures. For witness-guided BSE, preflight also attempts to confirm that execution under the witness reaches the named release function. The resulting diagnostics are returned to the planner for revision. An exact specification already recorded in $H$ is not dispatched again.

The extension ties witness bytes to the symbolic objects declared by an admitted target. Before release, the witness resolves supported input-dependent operations and the resulting constraints enter the path condition of the single state. At the release-function boundary, this state continues under KLEE's ordinary solver and search policy. When KLEE writes a KTest, it first obtains a satisfying assignment and retains witness bytes wherever they remain consistent with the final path condition. A final consistency check falls back to the solver assignment if the merged assignment does not satisfy that path condition.

For each program, we run \agolic in a Linux Docker container with 11 CPUs and 90~GiB of memory, using up to ten execution workers. Continuous KLEE starts 180 seconds before the first planning round and occupies one worker while it is active, leaving up to nine for planner-dispatched BSE. A BSE run receives 240 seconds by default, although its target specification may set a different limit, and the 3-hour deadline terminates any symbolic-execution process still active. The 180-second lead allows continuous KLEE to establish the initial coverage map before planning, while the 240-second default leaves time for successive feedback rounds within the three-hour limit; neither value is tuned per program. Each symbolic-execution process has 8~GiB of memory and a 10-second Z3 query limit \cite{z3solver}. The planner uses \texttt{gpt-5.4-2026-03-05} with medium reasoning and may take up to 40 agent turns in a round. When the deadline interrupts an active worker, the runtime recovers any tests already written for final corpus replay without determining the interrupted run's target reach or coverage gain.

\subsection{Benchmarks and Comparisons}
\label{sec:experiment-benchmarks}
\label{sec:experiment-comparisons}

We use seven C and C++ programs from the ConcoLLMic evaluation to compare \agolic with an LLM-guided concolic system \cite{luo2026agentic}. We leave out \texttt{libsoup} because its harness needs a network model that is unavailable in our container.

We run the released ConcoLLMic workflow and benchmark packages with \texttt{gpt-5.4} through a dedicated private OpenAI endpoint. Because requests through this endpoint have higher latency, we extend ConcoLLMic's published 30-minute coverage-stagnation window to 2 hours.

For each program, the continuous KLEE run gives us the tests produced without planner-dispatched BSE. For each of AFL++ \cite{fioraldi2020aflplusplus}, SymCC \cite{poeplau2020symcc} and SymSan \cite{chen2022symsan}, we generate three three-hour corpora per program using the benchmark packages released with ConcoLLMic and replay them through the native-input adapter. In these runs, AFL++ starts from the supplied benchmark seed corpus and receives no external dictionary.

\subsection{Coverage Measurement}
\label{sec:experiment-measurement}

Branch coverage depends in part on the files one decides to count. We therefore fix a source-file list for each benchmark, consisting of its harness and the project code exercised by that harness, including bundled parsers and decoders. Every corpus is replayed against the same native GCov build and measured over this list. No test input is removed, while headers, system libraries and unrelated programs in the same source tree remain outside the comparison. During each run of \agolic, the planner sees the GCov map $\mu_r$ for its configured source roots, produced by the round-level replay described in \autoref{sec:coverage}. Since the complete set of branches reachable through a harness is unknown, we report the number observed during concrete replay without treating the compiler-visible total as a reachability estimate.

Each corpus is replayed in its native format. Within \agolic, each replay performed during a planning round allows up to 10 seconds per input. For final measurement, generated KTests and materialised witnesses from \agolic are allowed up to 30 seconds per replay. Inputs from ConcoLLMic retain the 10-second limit in its released wrapper, while inputs from AFL++, SymCC and SymSan are allowed 1 second. Increasing the ConcoLLMic limit from 10 to 30 seconds leaves every covered branch set unchanged, while every AFL++, SymCC and SymSan input completes within 1 second. Branches observed during a completed replay count irrespective of the program's exit status, and the outcome and exact branch set remain in the replay record. An input that terminates before GCov writes its data contributes no branch. When enumerating inputs for replay, we exclude the \texttt{.ogg} and \texttt{.ttf} files produced by Oggenc and WOFF2 themselves. The reported \agolic branch sets are the union of the branches observed from generated KTests and materialised witnesses selected by the planner.

\section{Results}
\label{sec:results}

For each of the seven shared programs, we report the final corpus from one execution of \agolic and three runs of each external comparison system. Each execution of \agolic has a 3-hour deadline, though it may finish earlier under the conditions in \autoref{sec:orchestration}.

\subsection{Branch Coverage (RQ1)}
\label{sec:results-shared}

We begin with continuous KLEE, whose tests form part of the final \agolic corpus. The planned runs add coverage on every program. The arithmetic mean of the seven program-level ratios is $3.48\times$ and the median is $2.37\times$. The geometric mean is $2.88\times$, while pooling the branch counts across the seven programs gives $2.34\times$. The larger arithmetic mean reflects WOFF2's $10.09\times$ proportional gain from a continuous-run baseline of 99 branches; LibYAML has the largest absolute gain.

\begin{figure*}[!t]
\centering
\includegraphics[width=\textwidth]{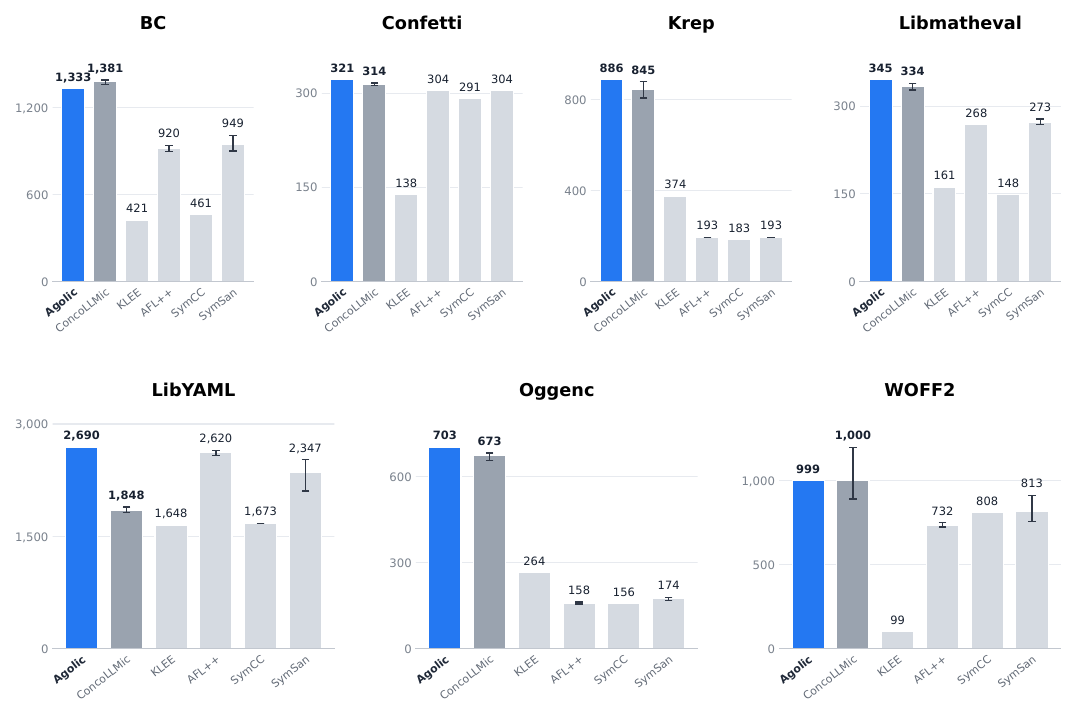}
\caption{Branches covered after concrete replay. Bars for repeated runs show the mean and whiskers show their range. Results within a panel use the same coverage build and source files; panels begin at zero and have independent scales. The \agolic corpus contains generated KTests and materialised witnesses.}
\label{fig:shared-tool-comparison}
\end{figure*}

\autoref{fig:shared-tool-comparison} gives the comparison with the other systems. \agolic covers more branches than any retained corpus produced by AFL++, SymCC or SymSan. Against ConcoLLMic, it exceeds all three runs on five programs.

Branch counts leave open whether the systems reach the same code, so \autoref{tab:branch-overlap} compares \agolic with the branches found across the ConcoLLMic repeats.

\begin{table}[!t]
\renewcommand{\arraystretch}{1.08}
\caption{Branches shared by \agolic and ConcoLLMic and branches found by only one system. The ConcoLLMic repeats are combined.}
\label{tab:branch-overlap}
\centering
\footnotesize
\begin{tabular}{@{}lrrr@{}}
\hline
\textbf{Program} & \textbf{Shared} & \textbf{\agolic{} only} & \textbf{ConcoLLMic only}\\
\hline
BC          & 1,256 & 77  & 231\\
Confetti    & 318   & 3   & 3\\
Krep        & 849   & 37  & 59\\
Libmatheval & 337   & 8   & 2\\
LibYAML     & 1,931 & 759 & 45\\
Oggenc      & 649   & 54  & 74\\
WOFF2       & 968   & 31  & 243\\
\hline
\end{tabular}
\end{table}

\agolic reaches branches absent from every ConcoLLMic run on all seven programs, including BC, where its total coverage is lower. LibYAML gives the strongest example, with 759 branches found only by \agolic, of which 702 lie in \texttt{scanner.c}, and 45 found only by the comparison system. Combining the AFL++, SymCC and SymSan runs gives a smaller branch set than \agolic on every program, with the Krep set contained entirely in \agolic. Even when all external runs are combined, \agolic still contributes branches on six programs; Confetti is the exception. The direct replay contribution of the witnesses is modest. Across the seven programs, materialised witnesses add 38 branches not already covered by generated KTests, while the KTests alone still cover $3.46\times$ as many branches as continuous KLEE on average. Removing those branches changes one comparison with ConcoLLMic: on Confetti, the KTests cover 311 branches, while the full \agolic corpus covers 321 and all three ConcoLLMic corpora lie between the two. We return to this concentration in \autoref{sec:discussion-concrete-prefix}, where we consider how the cost of reaching a program region and the choices left symbolic bear on the execution mode.

\subsection{The Planning Loop in Practice (RQ2)}
\label{sec:results-planning-loop}

Across these seven program evaluations, the planner completes 102 of 105 rounds and dispatches 411 BSE runs. All three failed planning sessions occur during BC and Oggenc; each is retried in a later round while continuous KLEE remains active.

The two modes are used in similar numbers overall, but the program-level choices are quite different. Krep uses Harness-Entry Mode throughout, whereas LibYAML uses only Witness-Guided Mode. \autoref{tab:planning-activity} gives the planning activity for each program and estimates the combined cost of planning and witness finding. Across these seven evaluations, more than 90\% of the recorded model input tokens were served from cache.

\begin{table}[!t]
\renewcommand{\arraystretch}{1.08}
\caption{Planning and BSE activity. Estimated costs include planning and witness finding; the total uses unrounded estimates.}
\label{tab:planning-activity}
\centering
\footnotesize
\setlength{\tabcolsep}{2.6pt}
\begin{tabular}{@{}lrrrrr@{}}
\hline
\textbf{Program} & \textbf{Rounds} & \textbf{H-Entry} & \textbf{W-Guided} & \textbf{KTests} & \shortstack{\textbf{Est.}\\\textbf{cost (\$)}}\\
\hline
BC            & 11 & 20 & 20 & 898   & 16.44\\
Confetti      & 13 & 17 & 38 & 630   & 13.61\\
Krep          & 19 & 80 & 0  & 1,993 & 34.89\\
Libmatheval   & 21 & 12 & 65 & 907   & 31.65\\
LibYAML       & 13 & 0  & 54 & 7,165 & 27.81\\
Oggenc        & 20 & 56 & 15 & 413   & 45.53\\
WOFF2         & 8  & 4  & 30 & 468   & 16.35\\
\hline
\textbf{Total} & \textbf{105} & \textbf{189} & \textbf{222} & \textbf{12,474} & \textbf{186.29}\\
\hline
\end{tabular}
\end{table}

One LibYAML planning trace shows how a later proposal revised an earlier configuration. After one specification used a three-byte file to target \texttt{yaml\_parser\_fetch\_next\_token}, the next round described the attempt as low-payoff and proposed four bytes on standard input for the same target, moving the release function from \texttt{yaml\_parser\_scan\_to\_next\_token} to \texttt{yaml\_parser\_fetch\_next\_token}.

To give a second view of this activity, \autoref{tab:execution-time} records the wall time consumed by the symbolic-execution processes. The continuous column records the lifetime of the accompanying continuous KLEE process, which spans the evaluation. The BSE column sums KLEE's recorded wall time across all 411 dispatched runs, including failed and deadline-interrupted runs. Together, the continuous and BSE runs account for 35.31 process-hours, approximately $1.83\times$ the time recorded for continuous symbolic execution alone. These figures exclude planning, target review, preflight, witness finding, concrete replay and other orchestration. They are process wall times rather than CPU times and, since BSE runs execute concurrently, their sum is not the elapsed duration of an evaluation.

\begin{table}[!t]
\renewcommand{\arraystretch}{1.08}
\caption{Recorded symbolic-execution process wall time. BSE time is summed across dispatched runs; combined time is therefore not elapsed evaluation time.}
\label{tab:execution-time}
\centering
\footnotesize
\begin{tabular}{@{}lrrr@{}}
\hline
\textbf{Program} &
\shortstack{\textbf{Continuous}\\\textbf{run (h)}} &
\shortstack{\textbf{BSE runs}\\\textbf{(h)}} &
\shortstack{\textbf{Combined}\\\textbf{(h)}}\\
\hline
BC            & 3.003 & 2.437 & 5.440\\
Confetti      & 1.803 & 2.488 & 4.291\\
Krep          & 2.965 & 1.479 & 4.444\\
Libmatheval   & 3.002 & 3.858 & 6.860\\
LibYAML       & 2.901 & 3.184 & 6.085\\
Oggenc        & 2.901 & 0.886 & 3.787\\
WOFF2         & 2.756 & 1.647 & 4.403\\
\hline
\textbf{Total} & \textbf{19.331} & \textbf{15.979} & \textbf{35.310}\\
\hline
\end{tabular}
\end{table}

Detailed outcome records are available for BC, Krep and WOFF2, so \autoref{tab:run-local-outcomes} reports their 154 dispatched BSE runs. Its first five rows refine the target-evidence classes in \autoref{tab:target-evidence} by also recording branch gains elsewhere; the final two rows have no such classification because replay evidence or a completed run was unavailable. Generated tests reach the named target for 110 runs; 85 of these add a branch somewhere and 25 add none. Another 34 runs do not produce a test that replays through the named target, although nine add coverage elsewhere. Replay evidence was unavailable for seven runs, so their target reach and coverage gain could not be determined. Three further BC runs were interrupted at the evaluation deadline before an outcome could be recorded; their 82 generated KTests remain in the final corpus.

\begin{table}[!t]
\renewcommand{\arraystretch}{1.08}
\caption{Recorded outcomes for the 154 BSE runs dispatched for BC, Krep and WOFF2.}
\label{tab:run-local-outcomes}
\centering
\footnotesize
\setlength{\tabcolsep}{4pt}
\begin{tabular}{@{}lr@{}}
\hline
\textbf{Outcome} & \textbf{Runs}\\
\hline
Target reached; target-local branch gain & 74\\
Target reached; branch gain only elsewhere & 11\\
Target reached; no branch gain & 25\\
Target not reached; branch gain elsewhere & 9\\
Target not reached; no branch gain & 25\\
Replay evidence unavailable & 7\\
Interrupted at the evaluation deadline & 3\\
\hline
\textbf{Total} & \textbf{154}\\
\hline
\end{tabular}
\end{table}

The witness finder is used in four of the seven program evaluations. Its 19 requests return 68 distinct inputs; the planner selects 48 and uses them in 83 dispatched target configurations, sometimes pairing an input with several targets or release functions. Eighty of these configurations use finder-supplied inputs in Witness-Guided Mode on LibYAML, Oggenc and WOFF2. Krep instead uses one finder-supplied file as a concrete argument in three Harness-Entry configurations. BC, Confetti and Libmatheval do not invoke the finder; their Witness-Guided configurations use inline values or existing workspace files.

\begin{figure}[!t]
\centering
\includegraphics[width=\columnwidth]{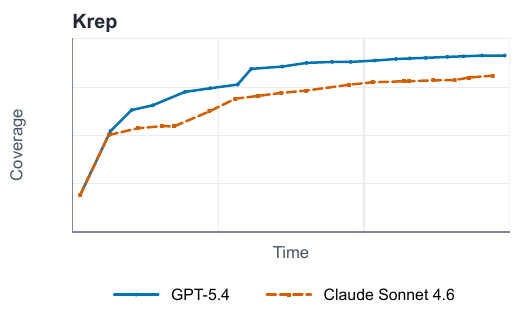}
\caption{Planner-visible coverage progression in one Krep execution with each planning model.}
\label{fig:model-planning-traces}
\end{figure}

We run \agolic on Krep again with Claude Sonnet 4.6. \autoref{fig:model-planning-traces} shows the planner-visible coverage progression in one execution with each planning model. We leave numerical scales off and read the traces qualitatively, since wall-clock progress depends on API latency and provider-side model optimisations. In these particular executions, GPT-5.4 reaches higher coverage, although one trace per model cannot tell us whether that difference would persist.

In the evaluated configuration, external reasoning chooses among the decisions made between runs, including the invocation, symbolic input surface and, when appropriate, a witness and release function. State selection during ordinary symbolic exploration remains with the symbolic execution tool. For 144 of the 154 BSE runs in \autoref{tab:run-local-outcomes}, replay determines both target reach and coverage gain. Of these runs, 94 generate at least one test that adds branch coverage during replay. These records let us see the planning loop in practice, with the planner repeatedly producing configurations whose runs yield tests with replay-confirmed coverage gains. We consider how the different outcomes can inform the next planning round in \autoref{sec:discussion-feedback}. Because \agolic also executes additional BSE runs, however, this evaluation cannot determine how much of the overall coverage difference comes from the planner's choices and how much comes from the additional symbolic execution, as discussed in \autoref{sec:threats}.

\section{Discussion}
\label{sec:discussion}

\agolic's present design grew out of two earlier attempts to place external reasoning within a symbolic execution run. Their limitations led us to plan between bounded runs, where a specification remains stable long enough for the resulting run to be executed and evaluated.

\subsection{Why KLEE}

We chose KLEE for reasons that are both practical and methodological. It operates on the LLVM-compiled C and C++ programs in the benchmark. Its POSIX runtime models the files, standard input and arguments used by their harnesses. KTests can also be replayed against native coverage builds. Since KLEE's source is available, we could add Witness-Guided Mode while retaining the search, constraint solving and test generation used during ordinary symbolic exploration.

One might reasonably ask why KLEE's existing seed support is not enough for this purpose. Under ordinary seeded execution, KLEE first advances states associated with the supplied input. The seed remains an assignment attached to symbolic states, however: KLEE still checks branch feasibility with the solver, may retain an unseeded sibling for the later search and may patch seed values when subsequent constraints conflict with them. The stricter \texttt{--only-replay-seeds} option discards unseeded states, but provides no chosen program point at which the state carrying the seed is handed to ordinary symbolic exploration. Concrete KTest replay takes the opposite course by fixing the input throughout execution, so it leaves no such hand-off either. Witness-Guided Mode supplies the missing transition by using witness values to establish one program state and releasing it into KLEE's ordinary exploration at the named release function, as described in \autoref{sec:witness-guided-bse}.

This choice also circumscribes the scope of our results. Symbolic PathFinder and angr use different languages, program representations and runtime interfaces \cite{pasareanu2010spf,shoshitaishvili2016angr}. We do not compare these symbolic execution tools or assume that planned BSE runs would behave in the same way. Our results apply only to this KLEE implementation.

\subsection{From Hard Frontiers to Planned Runs}

The ordinary KLEE runs used here begin symbolic exploration at the harness and do not use a witness during the initial execution. This leaves input-dependent choices before a region of interest to KLEE's search. In practice, path explosion and solver work can consume the run's budget before other uncovered code is reached. Our first response was to consult the agent where that cost had become visible in an individual state.

We called such a state a \emph{hard frontier}. Repeated windows of local stagnation and rising solver cost selected the state. KLEE continued its ordinary search while the agent examined a clone with its path condition, source context and nearby execution-tree nodes. The agent could propose values for symbolic bytes not yet fixed by the path condition. KLEE applied a proposal only when it remained feasible.

At first sight, the hard frontier offers a well-defined intervention point. The state is feasible, its constraints are known and KLEE has already shown that it is costly to advance. The same facts also limit the intervention. Any feasible patch must preserve every branch outcome already encoded in the state's path condition and can alter only the execution below that state. After resumption, states often terminated quickly or incurred further solver work in \texttt{memcpy} and \texttt{memset} with symbolic lengths or addresses. Patched states sometimes forked and explored additional paths. Over the 30--60 minute windows used at that stage, ordinary search often produced at least as much replayed branch coverage as the patched runs. A hard frontier may expose useful work within the selected path, but its earlier input-dependent branch outcomes remain fixed.

One development trial on libpng illustrated why the choice of symbolic object can matter as much as the value assigned to it. The harness exposed a 256-byte \texttt{data} object and a separate two-byte object encoding the logical \texttt{size}. Although the frontier heuristic nominated \texttt{data}, the proposed data patch was inconsistent with the path condition. The agent then supplied the feasible value 140 for \texttt{size}. The run recorded about $1.10\times$ as many covered instructions as each of three related trials, although concurrent ordinary search prevents attributing that difference to this proposal alone. The episode motivated exposing the symbolic input surface of a BSE run as a choice for the agent, beyond the bytes nominated by the frontier.

A more ambitious alternative was to place the agent inside KLEE's state-selection loop, where the model and symbolic execution tool operate on very different clocks. During one model turn, KLEE may select, fork and terminate many states, so a useful ranking had to return while the states it described still existed. A synchronous request would stall exploration, whereas an asynchronous request could describe a state set that had already changed. We therefore moved the decision outside KLEE's search loop. \agolic leaves the configured state-selection policy unchanged during ordinary symbolic exploration and asks the agent to configure BSE runs whose specifications remain stable long enough to execute and evaluate.

A run specification gives the planner a stable foothold because its invocation, symbolic input surface and release function still identify the same work when the model turn ends. \agolic uses an LLM to prepare these specifications, although a human could operate the same interface. The present study does not compare LLM and human planning.

\subsection{When a Concrete Prefix Fixes Too Much}
\label{sec:discussion-concrete-prefix}

A development trial on SQLite showed the other side of the same problem. The run had to reach \texttt{sqlite3CreateIndex} with useful input-dependent choices still symbolic. A \texttt{CREATE INDEX} statement first passes through the SQL parser. The parser both selects the statement type and constructs syntax and schema objects that the index-building code later consumes. By the time that code is reached, much of its behaviour is already determined by those objects.

Two earlier mechanisms supplied only part of what was needed. A fresh seeded run supplied the SQL bytes again, but KLEE still had to carry symbolic execution through the parser and none of the seeded trials emitted a test whose replay reached the routine within the allotted time. Path replay came closer because it could stop forcing recorded branch outcomes at the selected release function. Token values, addresses and allocation sizes nevertheless remained symbolic, and solver-query limits stopped execution before index construction. The release function alone was therefore not enough: execution also had to establish the state on the way there without incurring the symbolic work that had already defeated the seeded run. Witness-Guided Mode brings these requirements together by using witness values for supported input-dependent operations along the prefix and then releasing the resulting state. Yet a release at \texttt{sqlite3CreateIndex} inherits the statement choices already encoded in that state, so those earlier parser decisions cannot be revisited within that run.

Changing the solver was no panacea in these trials. STP failed while rewriting some large array constraints. Z3 allowed the trials to proceed further, although expensive parser and symbolic-memory queries remained. We therefore used Z3 with the per-query limit specified in \autoref{sec:experimental-setup}. This engineering choice applies to the present workload and does not support a general comparison of STP and Z3.

In these trials, SQLite offered no clean release boundary at which execution under the witness could end and ordinary symbolic exploration could begin. An early release preserves more symbolic choices and reintroduces path explosion and solver cost. A later release preserves the state needed by later code along with the decisions already made by the parser.

The concentration of \agolic-only LibYAML branches in \texttt{scanner.c}, reported in \autoref{sec:results-shared}, places much of the observed difference in scanner input processing, although it does not tell us why \agolic reaches those branches. Read alongside the SQLite trial, this directs attention to the cost of reaching a region, the state constructed on the way and the relevant choices that remain symbolic afterwards. The present evaluation does not establish whether these properties reliably predict which execution mode will be useful.

Witness-Guided Mode combines a concrete input with a planner-chosen release function. Corpus exchange is already established in hybrid testing, including Badger's exchange between fuzzing and bounded symbolic execution \cite{noller2018badger}. It remains to distinguish what comes from the input itself from what comes from the selected release point. A matched experiment that holds the concrete inputs fixed while varying how symbolic execution uses them would show what the target and release choices add beyond concrete-input reuse alone.

\subsection{Feedback Across Runs}
\label{sec:discussion-feedback}

The outcomes recorded in \autoref{tab:run-local-outcomes} do not all call for the same response, and even a run that adds no branch can inform the next planning round when the planner can distinguish what happened. Review and preflight expose a proposal that cannot be carried out, while concrete replay distinguishes a missed target from a reached target without a branch gain whenever replay evidence is available. Missing the target suggests revising the invocation, symbolic input surface, witness or release function. Reaching it confirms that the route works, while an absence of new coverage may mean that the relevant code is already covered, that a useful choice was fixed before release or that another worker arrived first.

A planning session may fail before producing an admitted target and be retried while continuous KLEE remains active. Tests produced by the continuous run before the later attempt can still add branches and change which contributions from BSE runs are new, making the continuous run both a fallback and a moving point of comparison.

\subsection{Coverage as the Search Objective}

Branch coverage serves two roles in this study. It gives every corpus a common outcome that can be checked independently through native replay, and its live map tells the agent what counts as progress. The reporting scope in \autoref{sec:experiment-measurement} defines the measured branch set, while the live source scope defines what the agent can see. Branch coverage does not tell us whether a generated test exposes a bug or reaches a worst-case execution. Making either the objective would change both the feedback shown to the agent and the runs it proposes. Studying such objectives would test how far run-level planning extends beyond coverage maximisation.

\section{Threats to Validity}
\label{sec:threats}

\paragraph{Internal validity}
The additional branches over continuous KLEE cannot be attributed to planning alone. A useful equal-capacity comparison must also decide how the additional workers differ; merely repeating the same configuration does not answer that question. Continuous KLEE uses one worker, whereas \agolic may use up to nine additional BSE workers. Since the BSE runs execute concurrently, the recorded contribution of one run also depends on whether another finds a branch first. The configurations in \autoref{sec:experimental-setup} and \autoref{sec:experiment-comparisons} use different resources and stopping rules. \agolic has a three-hour deadline, ConcoLLMic may stop after two hours without new coverage and the AFL++, SymCC and SymSan corpora are generated in three-hour runs. The latter are therefore duration-matched to \agolic's deadline, but the systems are not capacity-matched. The results compare final corpora under the stated configurations and do not provide equal-capacity measurements. Repeated runs with an equal-capacity, diversified KLEE portfolio are needed to separate the effect of the planner's choices from the effect of executing additional BSE runs.

\paragraph{Construct validity}
Branch totals depend on the coverage build and source files counted. We use the same native build and declared file list for every corpus within a benchmark, but another defensible list could produce different totals. The agent's live source roots need not match the reporting list, and we have not measured how this difference affects its choices. Each corpus format also requires its own replay adapter and per-input timeout, as described in \autoref{sec:experiment-measurement}. Increasing the timeout for ConcoLLMic inputs from 10 to 30 seconds changes no branch set, while every AFL++, SymCC and SymSan input completes within a second.

The reported \agolic corpus includes generated KTests and materialised witnesses selected by the planner. A witness may therefore contribute coverage through its own replay as well as guide execution before release. The direct contribution of witness replay is reported separately in \autoref{sec:results-shared}. Because the evaluated programs and related test corpora are public and the model's training data are undisclosed, we cannot determine whether prior familiarity with benchmark code or related inputs influenced planning. The finder's use of public web inputs is deliberate, but gives the evaluated configuration access to evidence not available to systems supplied only with fixed seed corpora. The results therefore characterise the complete configured system rather than a comparison in which all systems receive equivalent program knowledge; the curation costs of the supplied corpora and finder inputs are also unknown. Finally, branch coverage records control-flow reach. It does not show path difficulty, whether a bug was exposed or the semantic importance of a branch.

\paragraph{Conclusion validity}
For each program, we retain one final \agolic corpus. The agent's choices may vary from one execution to another and may produce different KLEE run configurations, so the results do not measure run-to-run variation or support statistical inference. The comparison systems have repeated runs, and the branch-identity comparison combines their branches before comparing them with one \agolic corpus. This gives those systems more opportunities to contribute a branch. The Krep plot contains one run per model and cannot rank the planning agents.

\paragraph{External validity}
The evaluation uses seven C and C++ programs from one previous study. It excludes \texttt{libsoup} because the required network model is unavailable and does not cover other languages, symbolic input models or application domains. For the seven-program comparison, we use the same KLEE implementation, solver settings and agent configuration throughout. The observed choices may behave differently with other harnesses, symbolic execution tools or planning agents, so broader claims require repeated evaluation in those settings.

\section{Conclusion}
\label{sec:conclusion}

We have shown how an external planner can use evidence from completed bounded runs to decide what a symbolic execution tool should attempt next, while leaving ordinary state exploration within each run to the tool itself. \agolic takes the bounded run as its unit of control, and in the branch-coverage configuration studied here an LLM-based agent uses source code, replayed coverage and earlier outcomes to choose later invocations, symbolic input surfaces and execution modes.

Across the seven C and C++ programs studied here, the final \agolic corpora extend the branch coverage of continuous KLEE on every program and cover more than three times as many branches on average. Each also covers more branches than any retained corpus produced by AFL++, SymCC or SymSan. Compared with ConcoLLMic, \agolic exceeds all three runs on five programs and reaches branches absent from all three comparison runs on every program, including BC, where its total is lower.

The choice of the run boundary is important because it gives the planner a stable specification to reason about and exposes choices that are already fixed once ordinary symbolic exploration begins. That specification can be revised for later exploration, while the symbolic execution tool continues to control the individual states during ordinary exploration within each run.

We have evaluated \agolic as a complete workflow, with the planner's choices and the additional BSE execution contributing to the same final corpus. Matched-capacity repetitions and ablations can distinguish their contributions and reveal which evidence and run configurations matter for different program regions. Applications to other symbolic execution tools and objectives can establish where the same division of labour remains useful beyond branch coverage.

For good reason, symbolic-execution research has devoted sustained attention to the search within a run. \agolic shows that important choices remain between runs, where the evidence already gathered can shape what the tool attempts next.

\bibliography{refs}

\end{document}